\documentclass[1,12]{elsarticle}

\usepackage{lineno,hyperref}
\usepackage{textcomp}
\usepackage{graphicx}
\usepackage{amsmath}
\usepackage{amssymb}
\usepackage{verbatim}
\usepackage{url}
\usepackage{gensymb}
\usepackage{booktabs}
\usepackage{setspace}
\usepackage{tabularx}
\usepackage{tabulary}
\usepackage{hhline}
\usepackage{multirow}
\usepackage{multicol}
\usepackage{float}
\usepackage{bm}
\usepackage{amsmath}
\usepackage{amssymb}
\usepackage{subfigure}
\usepackage{array}
\usepackage{caption}
\usepackage{color}

\journal{Materialia, accepted for publication}

\begin{document}

\begin{frontmatter}

\title{Criteria for slip transfer across grain and twin boundaries in pure Ni}


\author[mymainaddress,mysecondaryaddress]{E. Nieto-Valeiras}

\author[mymainaddress,mysecondaryaddress]{J. LLorca\corref{mycorrespondingauthor}}
\cortext[mycorrespondingauthor]{Corresponding author}
\ead{javier.llorca@upm.es , javier.llorca@imdea.org}

\address[mymainaddress]{IMDEA Materials Institute, C/Eric Kandel 2, 28906 – Getafe, Madrid, Spain}
\address[mysecondaryaddress]{Department of Materials Science, Polytechnic University of Madrid. E. T. S. de Ingenieros de Caminos, 28040 Madrid, Spain}

\begin{abstract}
Slip transfer at grain boundaries and annealing twin boundaries was studied in  polycrystalline Ni  by means of slip trace analysis. Slip transfer or  blocking was assessed in $>$ 200 boundaries and was related with geometrical criteria that establish the alignment between the active slip systems across the boundaries. It was found that slip transfer mainly occurs at low-angle regular grain boundaries and that a Luster-Morris parameter  $>$ 0.8 stands for the best criterion to assess slip transfer. In the case of coherent or incoherent  twin boundaries, slip transfer occurs when the residual Burgers vector is close to zero through dislocation cross-slip. 
\end{abstract}


\end{frontmatter}

\section{Introduction}
Grain boundaries (GB) play a major role on the mechanical behavior of polycrystalline materials due to the multiple interactions that can take place between dislocations and GB. In general, high angle GB are obstacles to the dislocation motion leading to the formation of dislocation pile-ups and to the well-known Hall-Petch effect \cite{EFN51, HSL18}. The stress concentrations induced by the pile-ups also facilitate the nucleation of damage by grain boundary cracking during fatigue \cite{JML22} while persistent slip bands parallel to the annealing twin boundaries (TB) are found to be dominant sites for fatigue crack nucleation in Ni-based superalloys \cite{Stinville2017, charpagne2021slip}. Nevertheless, there is ample experimental evidence that slip transfer can take place across GB when the slip systems in the neighbour grains are properly aligned  \cite{Shen1988, Lee1989} and different geometrical criteria were proposed in the literature to establish whether slip transfer will take place at a GB\cite{LC57, Lee1989, Luster1995, BMR16}. They are a function of the angle $\psi$ between the slip plane normals, the residual Burgers vector $\Delta b$ (which is directly related to the angle between slip directions $\kappa$ according to $\Delta b =2\sin (\kappa/2)b$) and the twist angle $\theta$ between the intersections of the slip planes with the GB. Nevertheless, the available experimental data were limited to a few observations in the transmission electron microscope \cite{BMR16} and it was not possible to discriminate the most accurate criteria. 

This limitation was overcome with the application of electron backscatter diffraction (EBSD) techniques in the scanning electron microscope (SEM) that, in combination with slip trace analysis, allows to determine whether slip transfer has taken place across a GB and the orientation of the slip systems in both grains across the boundary. Using this strategy, H\'emery {\it et al.} \cite{Hemery2018} identified $>300$ cases of slip transfer in a Ti-6Al-4V alloy and concluded that basal, prismatic and pyramidal slip systems seem to equally transfer to each other and that slip transfer is favored by low misorientation angles and high values of the Luster-Morris parameter $m' = \cos \psi \cos \kappa$. In another study in Al polycrystals, Bieler et al. \cite{Bieler2019} and Alizadeh et al. \cite{alizadeh2020criterion} also concluded that slip transfer is favored in low-angle GB and provided a geometrical criterion that indicated  that slip transfer is very likely to occur  when $m' > 0.9$ and the residual Burgers vector $\Delta b < 0.35b$. This criterion was recently introduced in a crystal plasticity model that takes into account the strengthening associated with the formation of dislocation pile-ups at GB when slip transfer is blocked \cite{Haouala2019} and the predictions of the Hall-Petch effect in various FCC polycrystalline metals were in better agreement with the experimental data when the possibility of slip transfer according to the previous geometrical criterion was included in the model \cite{nieto2021effect}.

Slip transfer in Ni-based superalloys polycrystals has been analyzed in several investigations. Abuzaid {\it et al.} \cite{Abuzaid2012} used a combination of EBSD and high resolution digital image correlation to capture the formation of pile-ups at GB (indicative of slip blocked) or the transmission of strains across the GB (indicative of slip transfer). This latter behavior was associated with small values of $\Delta b$, which seemed to be the best predictor of slip transfer, particularly for coherent TB where $\Delta b$ = 0 was associated with slip transfer via cross-slip. Gen\'ee {\it et al.} analyzed slip transfer in another Ni-based superalloy using EBSD and concluded that the necessary condition for slip transfer at coherent TB was given by $\theta$ = 0$^\circ$ (as opposed to the other possibility $\theta$ = 60$^\circ$). Regarding regular GB, they observed that slip transfer was favored by small values of $\psi$ and $\kappa$. They could not, however, provide a criteria because of the experimental scatter, which was attributed to the influence of the twist angle $\theta$ which cannot be determined from the EBSD information for regular GB. Finally,  Sperry {\it et al.} \cite{sperry2020slip} also combined EBSD with high resolution digital image correlation to study slip transfer in another Ni-based superalloy. They concluded that slip transfer in regular GB was associated with low misorientation angles and it was maximum when $m'$= 0.78 in twin boundaries.

These investigations on slip transfer in various Ni-based superalloys have not provided a quantitative geometrical criterion and sometimes have reached different conclusions, that may be due to the differences between the alloys used in each investigation. Moreover, the strengthening effect provided by the alloying elements in Ni-based superalloys hinders the development of slip bands in comparison to pure NI, where where larger datasets of slip blocking/transmission events can be obtained. Thus, we have analyzed slip transfer in more than 200 regular and twin boundaries of pure Ni foils by means of EBSD and slip trace analysis to rationalize the previous results and to provide robust geometrical criteria for slip transfer that can be used to predict slip transfer/blocking within the framework of crystal plasticity simulations.

\section{Material and experimental techniques}

\begin{figure}[t]
	\centering
	\includegraphics[width=\textwidth]{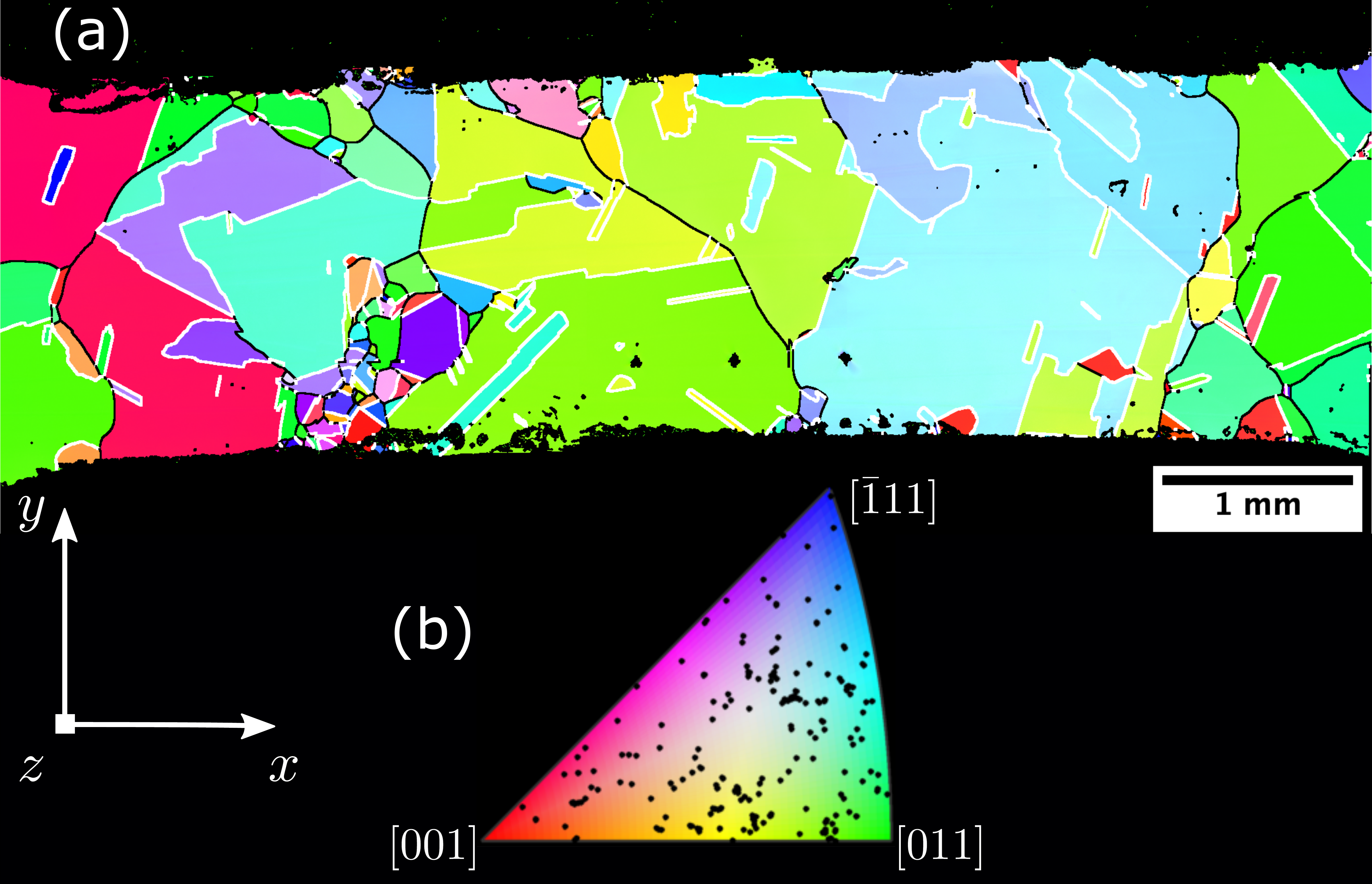}	
	\caption{(a) EBSD map of the surface of the dogbone specimen parallel to the rolling direction. The grain orientations with respect to the $z$ axis are indicated by the color, following the inverse pole figure. TB (which are associated with a misorientation angle of 60$^\circ$) are marked with white lines while regular GB are marked with black lines. (b) Inverse pole figure with respect to the $z$ axis. Each black dot corresponds to the crystallographic orientation of one grain in the dogbone specimen.}
	\label{ebsd_fig}
\end{figure}

Slip transfer across GB and TB was analyzed in a high purity (99.999\%) Ni foils of 1 mm in thickness manufactured by rolling, which were purchased from Goodfellow. Two dogbone micro-tensile specimens with a central gage section of $6\times2$ mm$^2$  were manufactured by electro-discharge machining with the tensile axis oriented at 0$^\circ$ and 45$^\circ$ from the foil rolling direction. The samples were encapsulated in quartz tubes filled with high-purity Ar (to prevent oxidation) and annealed at 800 $^\circ$C during 30 minutes in a tubular vacuum furnace to promote grain growth. EBSD maps of the gage section were obtained after metallographic preparation of the surface of the samples by manual grinding followed by electropolishing using a nitric acid-methanol solution (20\% volume of $\mathrm{HNO}_3$) at $-30$ $^\circ$C and 20 V. Four microindentations were placed in the central region of the gage to ease orientation and alignment of the samples inside the microscope (refer to Fig. \ref{ebsd_fig}). The EBSD maps revealed random texture and high presence of annealing twins in both samples and no further distinction is made between both specimens in this work. The average grain size was $240\pm250$ $\mu$m. The samples were tested under uniaxial tension in a Kammrath and Weiss micro-tensile testing machine up to 5\%  strain at a strain rate of $\approx 10^{-3}$ $s^{-1}$. Slip traces were visible in nearly all the grains  and the circular backscatter detector (CBS) was used in the SEM to obtain the optimum balance between brightness and contrast to identify  either slip transfer or blocking across GB. Slip transfer was analyzed in more than 200 GB, distributed in approximately half regular GB and half annealing TB.

\section{Results and discussion}

\begin{figure}[t!]
	\centering
	\includegraphics[width=\textwidth]{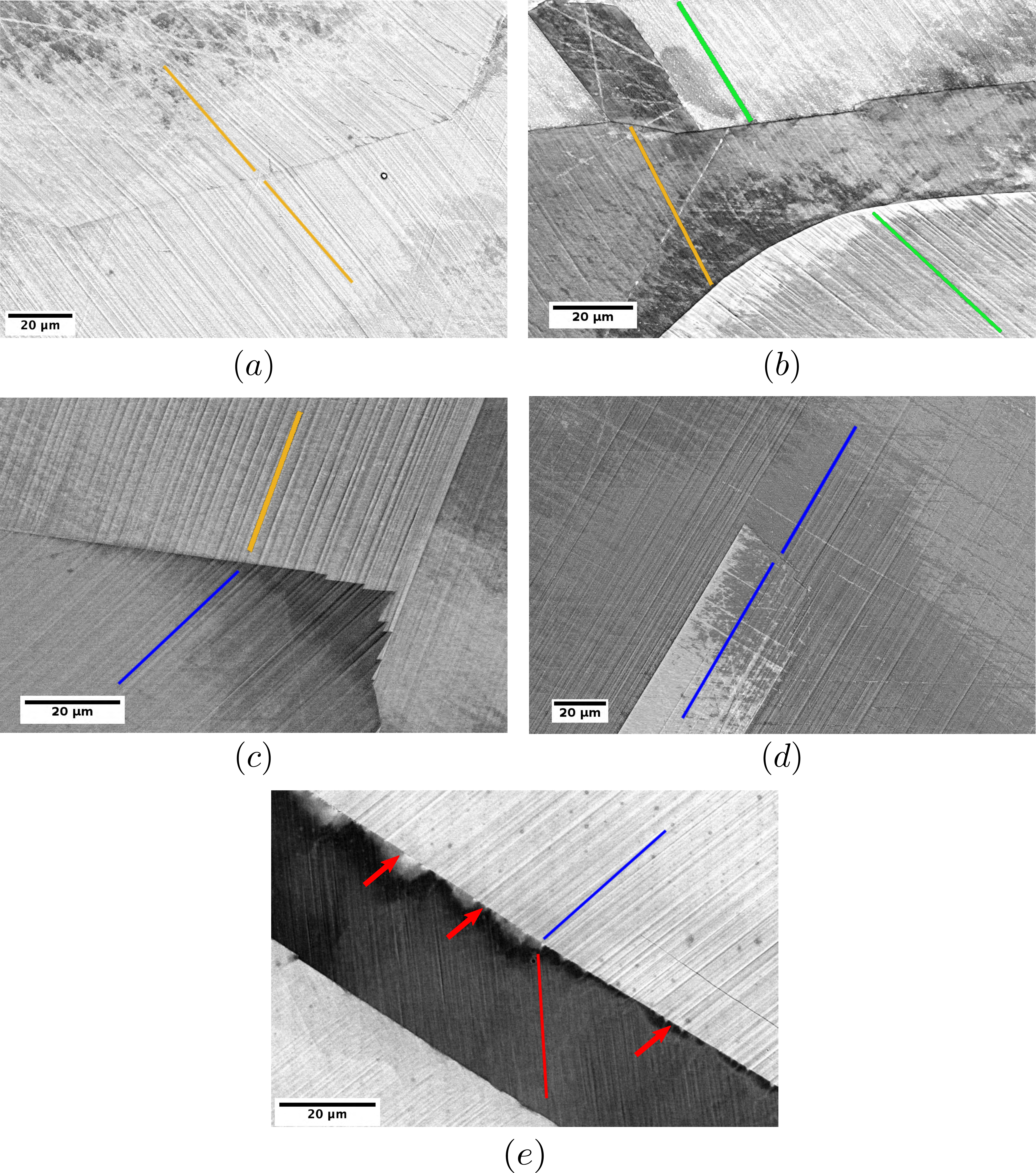}	
	\caption{Examples of slip transfer and slip blocking at different GB (a) Slip transfer across a regular GB. (b) Blocked slip at two regular GB, supported by the slip trace discontinuity and the presence of a ledge. (c) Slip transfer across a coherent TB. (d) Slip transfer across the end of the twin boundary (incoherent TB). The slip bands in the parent grain and in the twinned grain are parallel to each other and parallel to the twin boundary.  (d) Blocked slip at a TB supported by discontinuity between slip bands across the boundary and the formation of micro-volumes (marked by red arrows). The orientation of the trace of the active slip system in each grain according to slip trace analysis is marked by the lines in different colors.}
	\label{sem_micrographs}
\end{figure}

After deformation and inspection of the slip traces in each grain, GB were classified in two groups: boundaries where slip transfer was convincingly observed and boundaries where slip transfer was not observed. Convincing slip transfer was considered when the incoming and outgoing slip traces across a GB were continuous and clearly correlated while negligible topography along the GB was observed, indicating relatively homogeneous deformation between the grains (Figs. \ref{sem_micrographs}a, c and d). Furthermore, the large number of annealing twins promotes the activation of the slip systems parallel to the TB in the so-called parallel slip configuration near TB \cite{stinville2016combined,Stinville2017}, as observed in Fig. \ref{sem_micrographs}d. The annealing twins can span the whole parent grain beginning in one boundary and ending in the opposite one so the associated TB are fully coherent with the matrix. However, sometimes the annealing twin does not reach the opposite boundary, leading to the formation of another TB (perpendicular to the coherent TB) which is incoherent \cite{smallman2014chapter,pan2021interactions}. This is the case in Fig. \ref{sem_micrographs}d, in which the slip traces are parallel to the coherent TB (and, thus, neither slip transfer nor slip blocking occurs) and slip transfer is registered at the incoherent TB. Conversely, the slip blocking at the GB or TB was indicated by the lack of continuity between the observed slip bands across the boundary and this observation was reinforced by the existence of ledges in the boundary which is indicative of incompatibility of deformation between both grains (Fig. \ref{sem_micrographs}b). In some cases, the impingement of the slip traces on a GB leads to the formation of micro-volumes of deformation (Fig. \ref{sem_micrographs}e), which appear due to a high density of dislocations, local crystalline rotations and/or grain boundary shearing, and also indicate that slip transfer is not possible \cite{genee2017slip} .

The traces of the different \{111\} slip planes in each Ni grain on the specimen surface were obtained from a Matlab code from the crystallographic orientation of the grain provided by the EBSD scans obtained before deformation. It was assumed that the small lattice rotations induced by the applied deformation were negligible. To this end, the slip systems were rotated from the crystal to the sample reference frame using the orientation matrix computed from the Euler angles of the grain in Bunge convention. The trace orientation was computed from the cross product between the slip plane normal and the $z$ axis of the specimen in the sample reference frame.

The orientation of the traces for different slip systems were compared with the experimental slip traces (Fig. \ref{sem_micrographs}) to identify the active slip plane. Then, the Schmid factors for the possible slip directions in the active slip plane were determined assuming uniaxial tension from the stress tensor rotated to the crystal reference frame. From the three slip systems in the active slip plane, it was assumed that deformation occurred in the one with the highest Schmid factor, which is the most reasonable assumption \cite{zhao2008investigation,bieler2009role}. Once the two active slip systems across the GB or TB were identified, different metrics can be determined to assess whether they are able to predict slip transfer or blocking. Among them,  the Luster-Morris parameter $m'$ and the residual Burgers vector $\Delta b$ have been used previously \cite{Hemery2018, alizadeh2020criterion, Abuzaid2012, sperry2020slip} in metallic polycrystals. It should be noted that
If slip traces were only visible only in one grain across the boundary but not in the other, these GB or TB were not included in the analysis because it was not possible to identify the active slip plane in the latter grain.
 
\begin{figure}[t!]
	\centering
	\includegraphics[width=\textwidth]{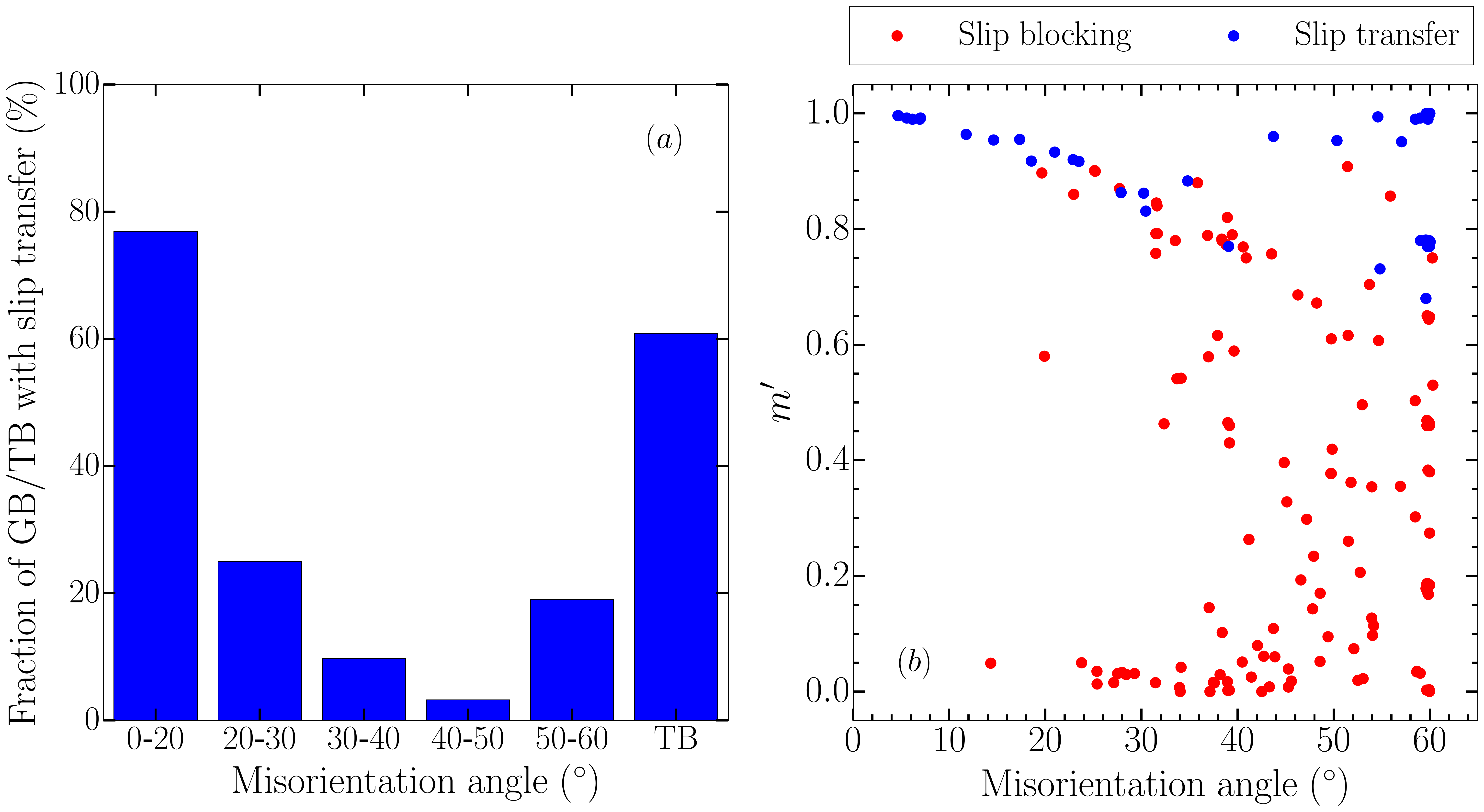}	
	\caption{(a) Effect of the misorientation angle on slip transfer for regular GB and TB. (b) GB and TB that present slip transfer (blue dots) and slip blocking (red dots) as a function of the Luster-Morris parameter $m'$ and the misorientation angle.}
	\label{mprime_mis}
\end{figure}

The analysis of slip transfer/blocking included 133 regular GB and 99 TB (75 coherent and 24 incoherent). They showed that slip transfer mainly took place at GB with low misorientation angle ($<$ 20$^\circ$) and in TB, which present a misorientation angle of 60$^\circ$ (Fig. \ref{mprime_mis}a). The experimental results of slip transfer/blocking at boundaries are plotted in Fig. \ref{mprime_mis}b as a function of the Luster-Morris parameter $m'$ and the misorientation angle. They show that slip transfer is strongly correlated with high values of $m'$ and that slip transfer in GB with high misorientation angle (in the range 20$^\circ$ to 60$^\circ$) was always associated to high $m'$, which indicates a good alignment of the slip systems. It should be noted, however, that slip blocking was observed in high angle GB when $m'$ was high. On the contrary, slip transfer was never observed when $m' < 0.65$. Thus, high $m'$ seems to be a necessary condition for slip transfer at regular GB but is not sufficient and this may be due to the influence of the twist angle $\theta$ that is unknown because it depends on the actual orientation of the GB throughout the sample. However, the limited number of instances of slip blocking with high $m'$ seems to indicate that the influence of $\theta$ is of second order as compared to $m'$.

Regarding annealing TB, slip transfer was observed in approximately 60\% and $m'$ was either equal to 1 (32\% out of the slip transfer instances across TB) and 0.77 (remaining 68\%) in these cases. $m'$ = 1 corresponds to slip transfer across an incoherent TB when the slip traces were parallel to the coherent TB (Fig. \ref{sem_micrographs}d) while $m'$ = 0.77 indicates that slip transfer took place at coherent TB by slip systems that were oblique the TB (Fig. \ref{sem_micrographs}c). The clustering of the slip transfer instances around these values of $m'$ are a result  of the limited number of combinations between incoming and outgoing slip systems across TB due to the 60$^\circ$ misorientation.

\begin{figure}[t!]
	\centering
	\includegraphics[width=\textwidth]{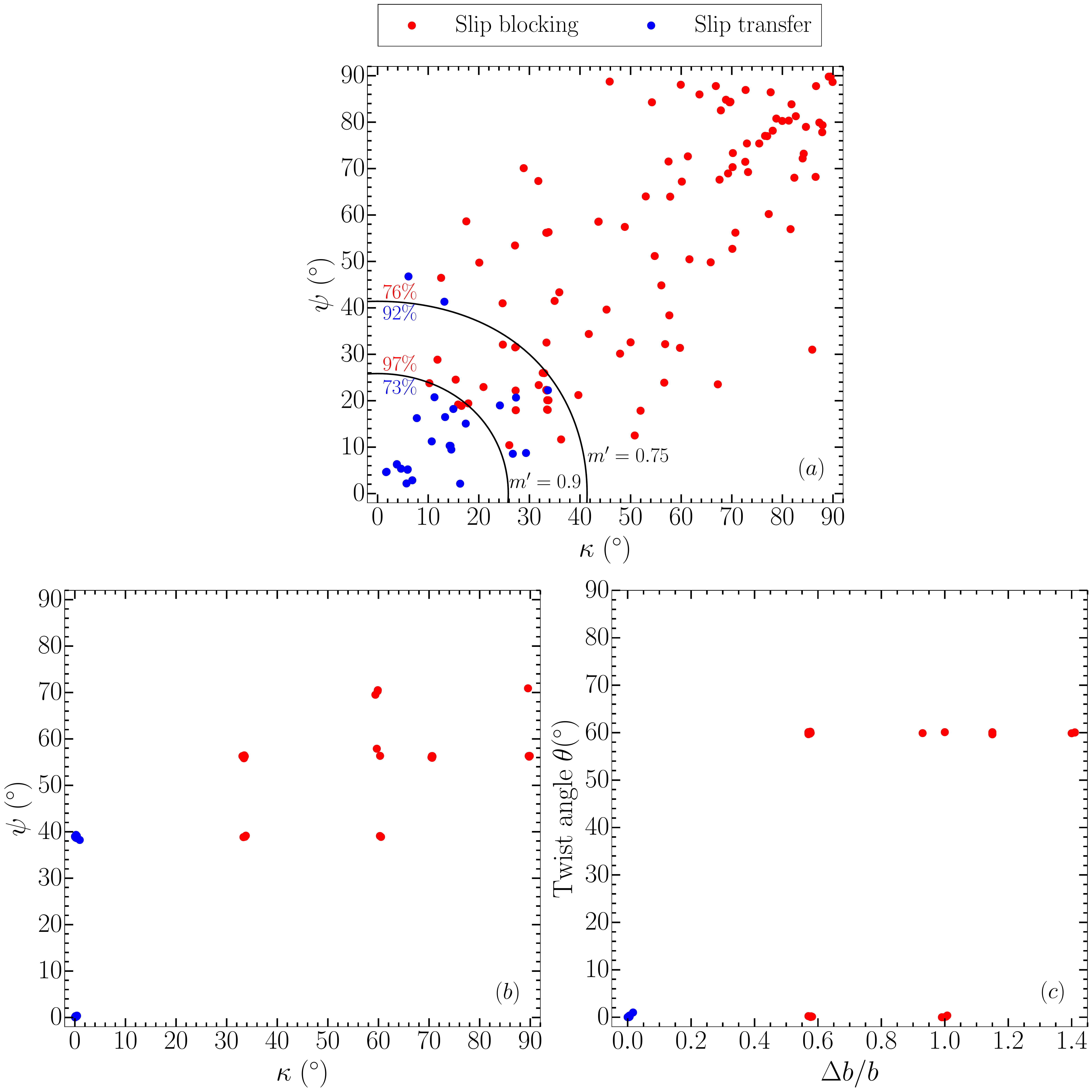}	
	\caption{Geometrical criteria for slip transfer in Ni. (a) Regular GB. (b) Coherent and incoherent TB. (c) Coherent TB.}
	\label{criteria_GBTB}
\end{figure}

The performance of $m'$ and $\Delta b$ as metrics  to predict slip transfer/blocking across GB and TB in Ni  is depicted in Fig. \ref{criteria_GBTB}. In the case of GB, the evidence of slip transfer and slip blocking in 133 GB is plotted in Fig. \ref{criteria_GBTB}a as a function of $\kappa$ (angle between slip directions of the incoming and out going slip planes, which is related to $\Delta b$) and $\psi$ (angle between slip plane normals of the incoming and outgoing slip planes). Practically all the slip transfer instances are found in the lower left corner of the figure, indicating that high values of $m'$ is an excellent predictor of slip transfer. For instance, the criterion of $m' \ge$ 0.9 for slip transfer (represented by a circular line in Fig. \ref{criteria_GBTB}a) predicts correctly 73\% and 97\%, respectively, of the slip transfer and slip blocking events. if $m' \ge$ 0.75 is chosen as the slip transfer criterion, 92\% and 76\% of slip transfer and slip blocking events are predicted correctly, respectively.  Thus, $m' >$ 0.8 seems to be an accurate criterion for slip transfer across regular GB in Ni, similar to the results reported previously for Al \cite{alizadeh2020criterion}. The few experimental data of slip transfer/blocking that do not fulfill this criterion may be attributed to GB with very large twist angle $\theta$, which impedes the slip transfer from one grain to the neighbor but this factor seems to be of second order as compared with the role played by $m'$.

In the case of coherent and incoherent TB (Fig. \ref{criteria_GBTB}b), slip transfer seems to be controlled by the angle between slip directions $\kappa$ or, in other words, by the residual Burgers vector $\Delta b$. Slip transfer was found in 43 coherent ($\psi \approx 39 ^\circ$) and 21 incoherent TB ($\psi \approx 0 ^\circ$) when the $\Delta b \approx 0$ and slip blocking was found in 35 coherent and incoherent TB when $\Delta b > 0$. When $\psi$ and $\kappa$ were close to zero, slip transfer took place across the incoherent TB present in unfinished annealing twins by the nearly perfect alignment of the active slip systems in the parent and in the twinned regions (Fig. \ref{sem_micrographs}d). When the angle between slip plane normals $\psi$ was not zero, slip transfer was only observed when $\kappa$ (and, thus, $\Delta b$) were close to zero. Hence, slip transfer was independent of the angle between slip system normals $\psi$, indicating that the incoming and outgoing slip systems may have very different orientations. Obviously, this abrupt change in the slip plane orientation, together with a negligible residual Burgers vector is indicative of slip transfer by dislocation cross-slip at TB  \cite{Abuzaid2012, ESP20}. In addition, the role played by the twist angle $\theta$ on slip transfer can be analyzed in the case of coherent TB because the orientation of the boundary between parent and twin grain is perfectly defined from the EBSD analysis. This angle was calculated as the angle between the projections of the slip plane traces on the twin boundary plane as indicated by Zhai {\it et al.} \cite{zhai2000crystallographic}.
\begin{equation}
\theta=\arccos(\bm{n}_{TB}\times\bm{n}_\alpha\cdot \bm{n}_{TB}\times\bm{n}_\beta)
\end{equation}
\noindent where $\bm{n}_{TB}$ is the normal to the coherent TB and $\bm{n_\alpha}$ and $\bm{n_\beta}$ stand for the slip plane normals to the incoming and outgoing slip systems, respectively.
As shown in Fig. \ref{criteria_GBTB}c, only two twist angles, $\theta$ = 0$^\circ$ and 60$^\circ$ can be found in these coherent TB. Slip blocking was found in the 24 TB with $\theta$ = 60$^\circ$ while 43 of the 51 TB with $\theta$ = 0$^\circ$ showed slip transfer. These are the ones whose residual Burgers vector is close to 0, indicating that small values of $\Delta b$ stand for the necessary condition for slip transfer in annealing TB in Ni and cross-slip is the physical mechanism responsible for slip transfer.

\section{Conclusions}

 The analysis of slip transfer and blocking in more than 200 GB in polycrystalline Ni samples has shown that slip transfer is likely to occur across low-angle regular GB and annealing TB. 
In the case of regular GB, slip transfer is very likely to occur if the Luster-Morris parameter $m' >$ 0.8 and it is very unlikely to occur otherwise. Slip transfer at coherent and incoherent TB is controlled by the residual Burgers vector and takes place when $\Delta b \approx 0$ regardless of the angle between slip normals and of the twist angle between the incoming and outgoing slip planes. In this latter case, slip transfer seems to be controlled by dislocation cross-slip. These geometrical criteria can be easily implemented into crystal plasticity models \cite{Haouala2019}  and open the way to predict the mechanical response of polycrystals taking into account the effect of slip transfer/blocking on the strength and failure mechanisms at GB \cite{nieto2021effect}.

\section*{Declaration of Competing Interests}

The authors declare that they have no known competing financial interests or personal relationships that could have appeared to influence the work reported in this paper.

\section*{Acknowledgments}

This investigation was supported by the European Research Council (ERC) under the European Union's Horizon 2020 research and innovation programme (Advanced Grant VIRMETAL, grant agreement No. 669141) and by the HexaGB project of the Spanish Ministry of Science and Innovation (reference RTI2018-098245, MCIN/AEI/ 10.13039/501100011033). 

\section*{Data availability}

The original data of this study can be obtained by request to the corresponding author.


\end{document}